\documentclass[a4paper]{jpconf}
\usepackage{graphicx}
\usepackage{amsmath}
\usepackage{subfigure}

\begin{document}
\title{$\Delta\eta - \Delta\phi$ correlations and the ridge structure in STAR}

\author{L. C. De Silva for the STAR collaboration}

\address{The Department of Physics, University of Houston, 617 SR Building 1, Houston, Texas 77204-500}

\ead{desilva@uh.edu}

\begin{abstract}
Triggered di-hadron correlation studies using central Au+Au collisions at $\sqrt{S_{NN}}$ = 200 GeV in STAR revealed a novel Òridge-likeÓ structure in two dimensions ($\Delta\eta$, $\Delta\phi$) for high $p_{T}$ particles. A similar structure is also present in an inclusive un-triggered di-hadron correlation analysis. We study the $\langle p_{T}\rangle$ evolution of di-hadron correlations by increasing the lower $p_{T}$ acceptance of both charged particles. A smooth evolution of data is observed and our results reproduce the triggered analysis structure near $\langle p_{T}\rangle$ = 2.7 GeV/c. We quantify the correlation structure evolution by fitting a model function. The model function emphasizes possible initial state fluctuation contributions via the use of higher harmonic model components $v_{n}$ (n=1,2,3,4,5) and a remainder which is modeled via an asymmetric 2d Gaussian. The extracted parameters are compared to model predictions and p+p data at $\sqrt{S_{NN}}$ = 200 GeV and possible origins of the nearside structure are discussed.\end{abstract}

\section{Introduction}
Two dimensional, high $p_{T}$ di-hadron correlation studies in STAR revealed a novel correlation structure elongated in pseudo-rapidity ($\Delta\eta$) [1]. This structure which is known as the "ridge" was observed by applying an asymmetric transverse momentum cut on trigger (4 GeV/c $>$ $p_{T}$ $>$ 3 GeV/c) and associate ($p_{T} > 2 GeV/c$) particles [2]. Furthermore, it was revealed that the structure persists up to trigger particle $p_{T}$ $>$ 6 GeV/c, possibly suggesting a jet-related origin.
\\
An independent analysis [3] also observed the ridge correlation for charged particles with $p_{T}$ $>$ 0.15 GeV/c for both trigger and associate particles. This analysis differs from the above mentioned one in three key aspects. It uses all possible correlated particle pairs, it applies a common lower $p_{T}$ threshold to all particles and the correlation function is normalized in a way that yields correlated particles per final state charged particle. The measured structure is not attributed to expected physical processes and in the past several attempts have been made to explain the long-range correlations through the medium modification of elementary processes such as jets or flux tube formation [4,5,6]. 
\\
For our study we adopt the un-triggered analysis method and focus on the correlation structure in the 0-10$\%$ centrality bin where we have the highest probability of forming a deconfined medium of quarks and gluons. We then investigate the evolution of the correlation structure as a function of transverse momentum. The structure is studied by increasing the lower transverse momentum acceptance for both particles. The resulting correlation is modeled via an empirical model fit function. We discuss our findings in the context of models that describe medium properties and modified jet phenomena in deconfined QCD matter.
  
\section{Data and analysis}
The data used in this analysis were collected during two runs, in 2004 and 2009, using the STAR detector at the Relativistic Heavy Ion Collider (RHIC) Brookhaven National Lab (BNL), Long Island, New York. For the results presented here we have analyzed 11 million Au+Au central trigger events and 264 million p+p minimum bias events at $\sqrt{S_{NN}}$ = 200 GeV. Only events with a primary vertex within $|$x,y$|$ $<$ 3cm and $|$z$|$ $<$ 25cm were taken into account. Charged particle tracks were reconstructed by the STAR TPC only and constrained to the primary vertex. The correlation function was calculated only for tracks within the kinematical cuts of $|$$\eta$$|$ $<$ 1.0, full 2$\pi$ acceptance in azimuth $\phi$ and $p_{T}$ = 0.15-15 GeV/c. The correlation measure for this analysis is defined as follows:

\begin{align}
          \frac{\Delta\rho}{\sqrt{\rho_{ref}}} &= \frac{\rho_{sib}\--\rho_{ref}}{\sqrt{\rho_{ref}}}
\end{align}                            
                                             
\begin{figure}[h]
\centering
\includegraphics[width=1.0\textwidth]{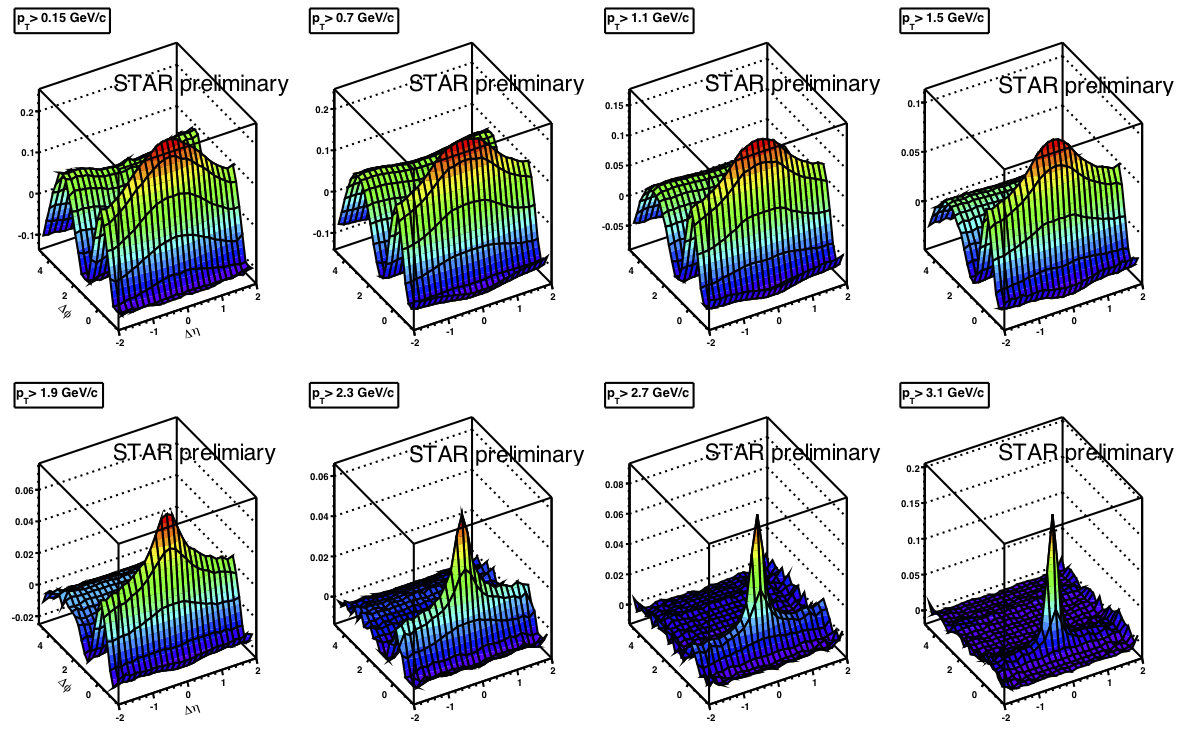}
\caption{Evolution of the raw normalized correlation structure for selected cuts on $p_{T}$ of the particles (increasing $\langle p_{T}\rangle$ from left to right) at 0$\--$10$\%$ centrality bin.}
\label{figure:one}
\end{figure}

The symbol $\rho_{sib}$ refers to the total correlated and uncorrelated pair density as a function of $\Delta\eta$ and $\Delta\phi$, for hadrons from a single event. $\rho_{ref}$ refers to the number of uncorrelated pairs as a function of $\Delta\eta$ and $\Delta\phi$, for hadrons from different events. $\Delta\rho$ is therefore the number of correlated pairs in a single event. The denominator is the square root of the number of background pairs, which corresponds to the number of particles. The correlation function therefore measures the number of correlated pairs per particle. In order to correct for acceptance effects, we sub-bin the z vertex into ten 5cm bins and allow a multiplicity window of 50 for event mixing. The conversion electron/positron background is reduced by using a 1.5$\sigma$ dE/dx cut on electrons in the momentum ranges, 0.2 $<$ $p_{T}$ $<$ 0.45 GeV/c and 0.7 $<$ $p_{T} $ $<$ 0.8 GeV/c. In order to analyze the transverse momentum evolution of di-hadron correlations in the 0-10$\%$ centrality bin we steadily increase the lower particle $p_{T}$ to gain further insight into the origin of the long range correlations. The correlation function evolution for eight selected bins is shown in Fig.~\ref{figure:one}. We model structures via the use of an empirical function and infer potential underlying physics mechanisms of the near side long range correlation. In the following section we discuss the model fit function.

\section{Model function study}
The main part of the model function is based on recent theory developments related to initial energy density fluctuation contributions to the two particle correlation structure [7]. It has been pointed out that the fluctuation contributions are not negligible even in the raw data spectrum and should be observable in very central events using high $p_{T}$ charged particles, as shown in Fig.~\ref{figure:two}. The remaining structure after subtracting out the higher order Fourier harmonics, which we term the remainder, is modeled via an asymmetric 2d Gaussian, and thus can be attributed to modified jet phenomena. The remainder properties are quantified by comparing to p+p data.   

\subsection{Evidence for initial density fluctuations in data}
\begin{figure}[h]
\centering
\includegraphics[width=1.0\textwidth]{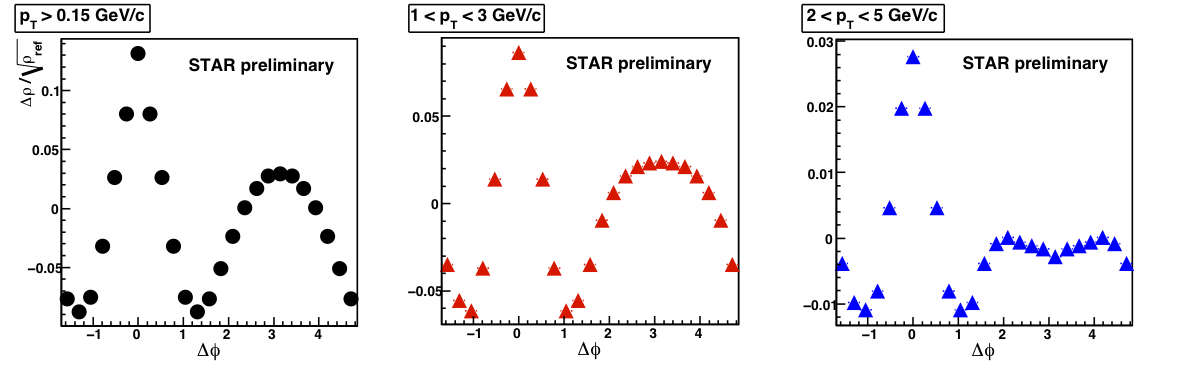}
\caption{Angular correlations as a function of $p_{T}$ cuts on the pairs for 0-1\% central data.}
\label{figure:two}
\end{figure}

Following the theory prediction [7], we focus our attention to the 0-1$\%$ centrality bin. Fig.~\ref{figure:two} shows  the projected 2d correlation structure on to $\Delta\phi$ as a function of $p_{T}$ revealing a double hump structure in the 2 $<$ $p_{T}$ $<$ 5 GeV/c interval for the away side correlations. This observation is consistent with the kinematic ranges predicted by theory and most likely due to triangular flow [8].

\subsection{Model function}
We initially attempt to describe the data using a maximum number of Fourier coefficients, which for our data requires coefficients up to n = 5 in the cos(n$\Delta\phi$) series. However the remaining structure reveals a significant residual, thus requiring an additional asymmetric 2d Gaussian term. The complete function can be broken down as follows.

\begin{eqnarray*}
  F &=&  c_{0}\; \qquad \qquad \qquad \qquad \qquad \qquad \qquad \; \; \; \; \; \: \propto \textit{Offset ($f_{1}$)}\\
        && +\, c_{1}\ast cos(1\Delta\phi)\; \qquad \qquad \qquad \qquad \qquad \: \propto \textit{Away-side momentum conservation ($f_{2}$)}\\ 
        && +\, c_{2}\ast cos(2\Delta\phi)\; \qquad \qquad \qquad \qquad \qquad \: \propto \textit{Second order Fourier harmonic ($f_{3}$)}\\ 
        && +\, c_{3}\ast cos(3\Delta\phi)\; \qquad \qquad \qquad \qquad \qquad \: \propto \textit{Third order Fourier harmonic ($f_{4}$)}\\ 
        && +\, c_{4}\ast cos(4\Delta\phi)\; \qquad \qquad \qquad \qquad \qquad \: \propto \textit{Fourth order Fourier harmonic ($f_{5}$)}\\
        && +\, c_{5}\ast cos(5\Delta\phi)\; \qquad \qquad \qquad \qquad \qquad \: \propto \textit{Fifth order Fourier harmonic ($f_{6}$)}\\
        && +\, c_{6}\ast exp(-0.5\ast((\Delta\phi/c_{7})^{2} + (\Delta\eta/c_{8})^{2})) \; \; \: \propto \textit{Modified jet fragmentation ($f_{7}$)}
\end{eqnarray*}

\begin{eqnarray*}
F &=& \textit{$f_{1} + f_{2} + f_{3} + f_{4} + f_{5} + f_{6} + f_{7}$}
\end{eqnarray*}

In the following section we discuss and interpret the model parameters we extract.

\section{Results and Discussion}
\subsection{Results}
 \begin{figure}[h]
        \centering
        \subfigure[The remainder evolution as a function of $p_{T}$]
       {
        \includegraphics[width=0.9\textwidth]{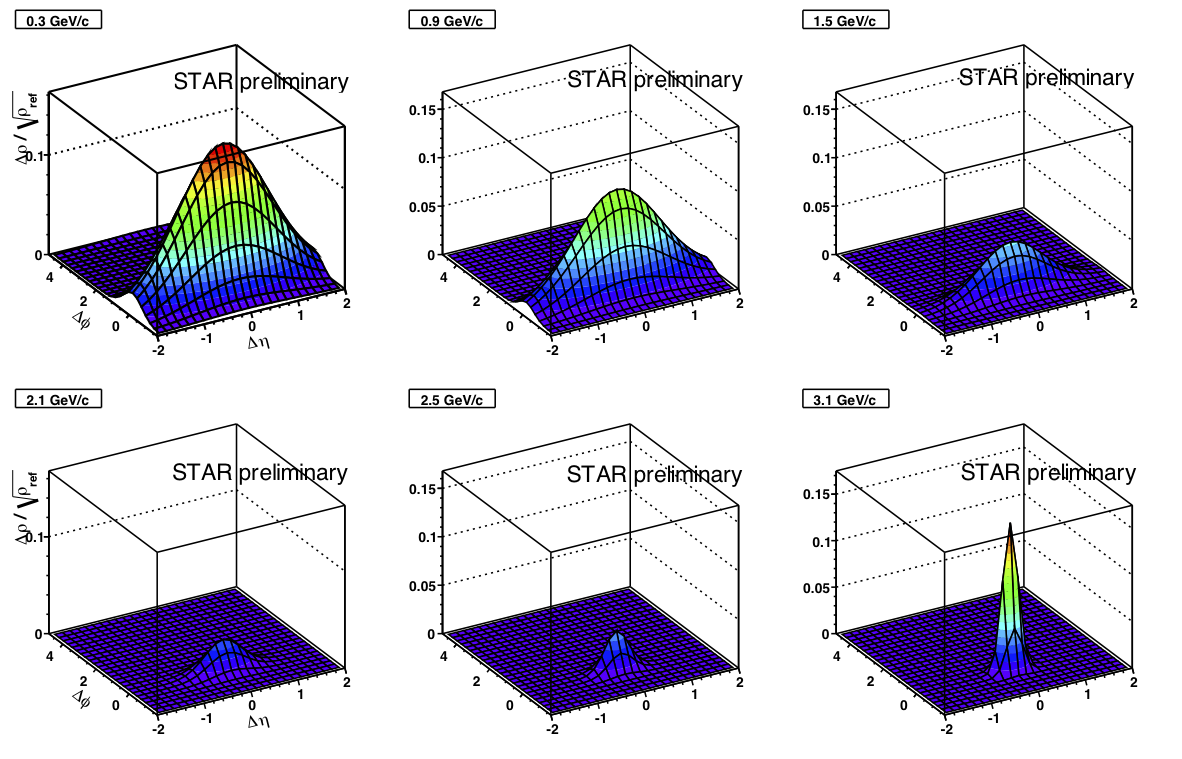}
        \label{figure:3a}
       }
       \subfigure[$v_{2} + v_{3} + v_{4} + v_{5}$ evolution as a function of $p_{T}$]
      {
       \includegraphics[width=0.9\textwidth]{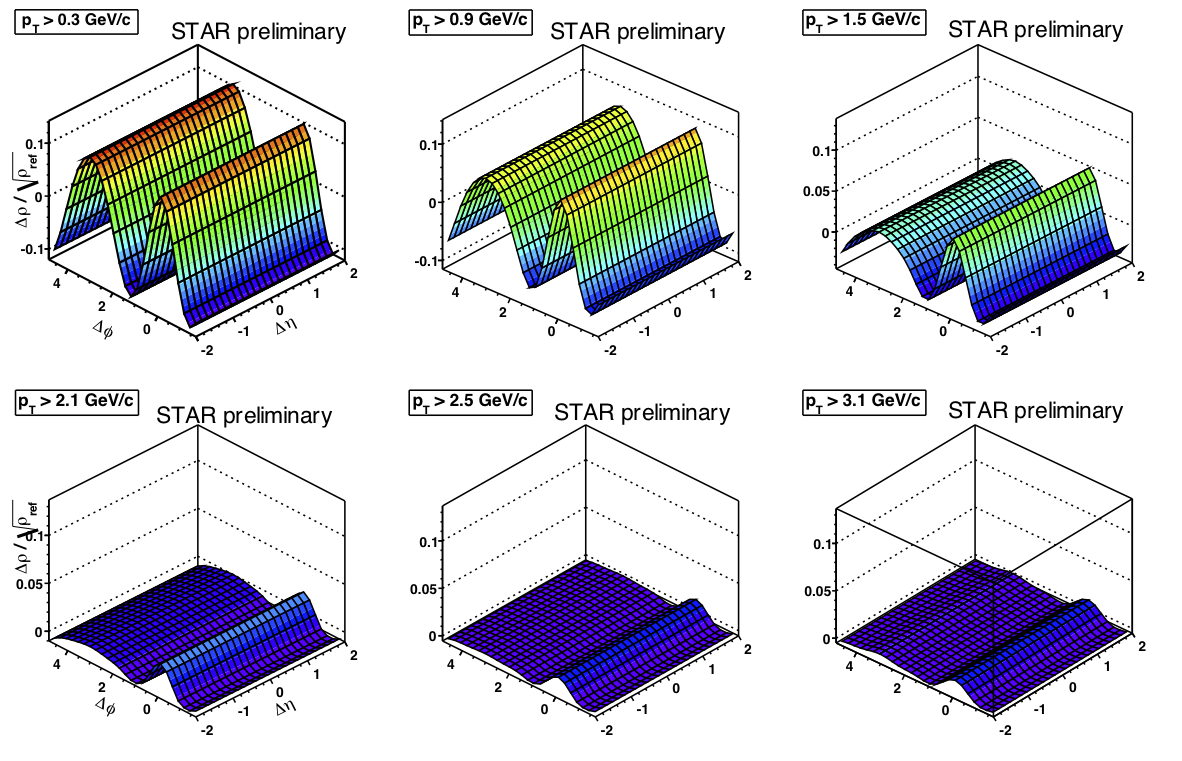}
       \label{figure:3b}
      }
      \\
      \caption{The evolution of the "remainder" (a) and the summed $v_{n}$ structure (n = 2 - 5) (b), as a function of the lower cut on $p_{T}$ of the particles.}
      \label{figure:three}
\end{figure} 

In this paper we mainly focus on the evolution of the model components shown in Fig.~\ref{figure:three}. Fig.~\ref{figure:3a} shows the evolution of the remainder component after subtracting out the harmonics (n=1-5) contributions to the correlation structure. Fig.~\ref{figure:3b} shows the sum of all harmonics (n=2-5) contributions. The remainder parameters might shed some light on potential modification to jets in the medium whereas the harmonics parameters can be used to test predictions from hydrodynamical models.\\
 
\begin{figure}[ht]
\centering
\includegraphics[width=0.9\textwidth]{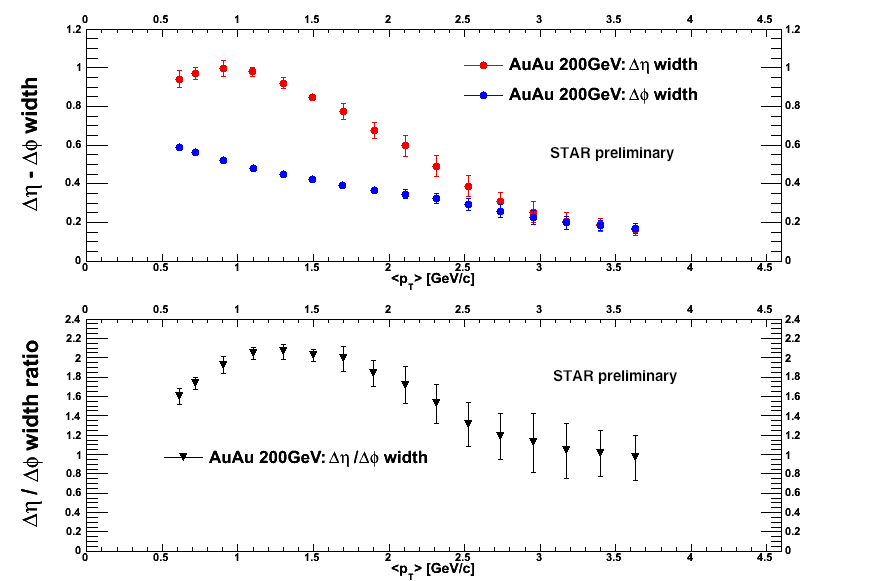}
\caption{The $\langle p_{T}\rangle$ evolution of $\Delta\eta$, $\Delta\phi$ widths of the remainder.}
\label{figure:four}
\end{figure}

The $\Delta\eta$-$\Delta\phi$ asymmetry of the remainder widths are shown in Fig.~\ref{figure:four}. We observe an asymmetry below $\langle p_{T}\rangle$ $\approx$ 2.5 GeV/c. The largest asymmetry is observed at $p_{T}$ $\approx$ 1.3 GeV/c where the $\Delta\eta$ width is a factor of two greater than the $\Delta\phi$ width.

\begin{figure}[ht]
        \centering
        \subfigure[$\Delta\phi$ width comparison]
       {
        \includegraphics[width=0.47\textwidth]{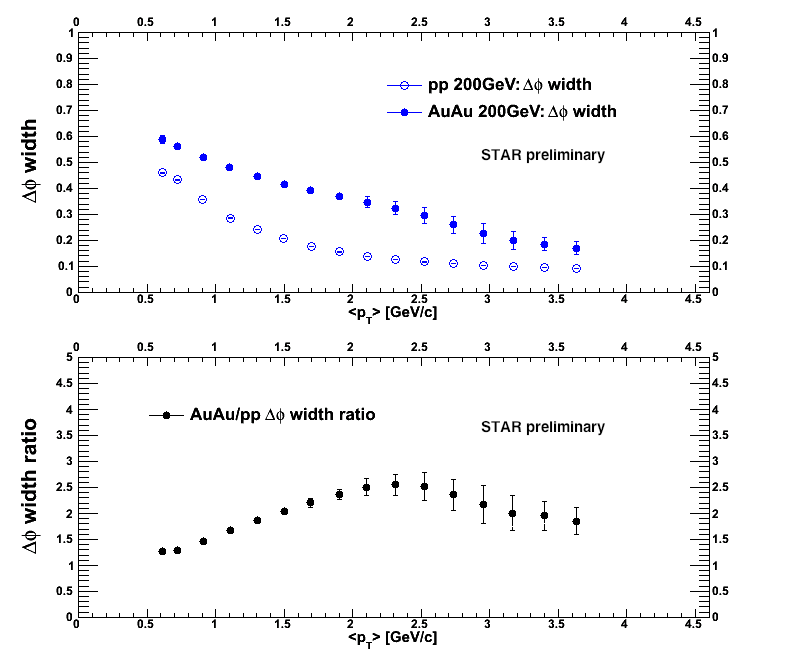}
        \label{figure:5a}
       }
       \subfigure[$\Delta\eta$ width comparison]
      {
       \includegraphics[width=0.49\textwidth]{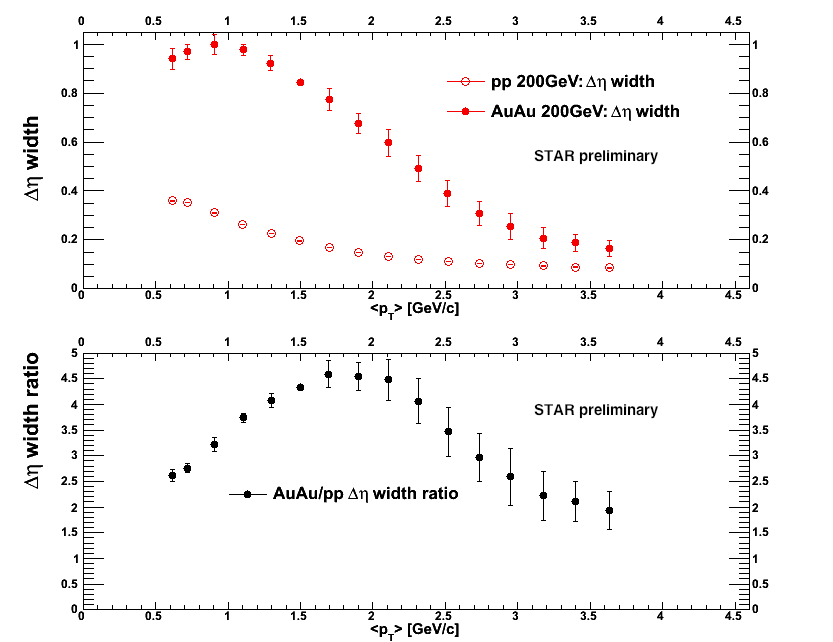}
       \label{figure:5b}
      }
      \caption{Comparison of the remainder $\Delta\eta$, $\Delta\phi$ widths to corresponding p+p widths.}
      \label{figure:five}
\end{figure}

We compare the extracted parameters to the same parameters from our analysis of p+p data in Fig.~\ref{figure:five}. In the region above 2.5 GeV/c where the $\Delta\eta$-$\Delta\phi$ widths are symmetric in Au+Au, the widths are about a factor of two larger than in p+p. Below that $\langle p_{T}\rangle$ value, we observe modification in both $\Delta\eta$ and $\Delta\phi$ remainder withs compared to corresponding p+p widths. Furthermore, the maximum relative width modulation (AA/pp) is a factor of two greater in $\Delta\eta$ width compared to $\Delta\phi$ width with the maximum modulations occuring at two different $\langle p_{T}\rangle$ values.

\begin{figure}[ht]
\centering
\includegraphics[width=0.5\textwidth]{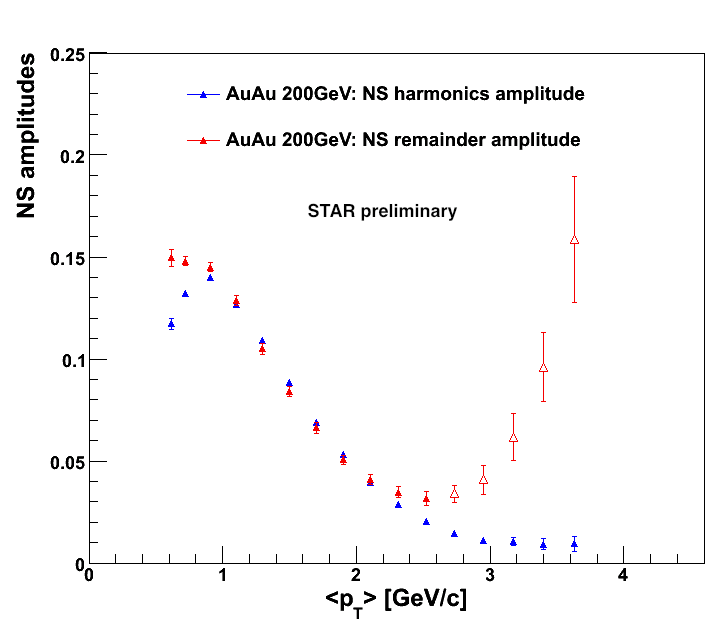}
\caption{Remainder amplitude variation compared to summed harmonic ($v_{2}+v_{3}+v_{4}+v_{5}$) amplitude.}
\label{figure:six}
\end{figure}     

Fig.~\ref{figure:six} shows the comparison between the remainder and the sum of all harmonics amplitudes. The amplitudes are comparable in the 0.9 $<$ $p_{T}$ $<$ 2.1 GeV/c range. The summed harmonics amplitude, which extracts the near-side long range correlation in our model, saturates at a small value at high $p_{T}$. The correlation strength of the added harmonic structure peaks at a $\langle p_{T}\rangle$ $\approx$ 0.9 GeV/c which is higher than the $\langle p_{T}\rangle$ for bulk particles thus indicating that long-range correlations might carry particles coming from jet fragmentation (hard processes). The asymmetric remainder amplitude drops with $\langle p_{T}\rangle$ and the symmetric remainder amplitude rises. The behavior of the symmetric remainder amplitude at $p_{T}$ $>$ 2.5 GeV/c can be understood based on more correlated jet fragment pairs in a jet like cone.

\begin{figure}[ht]
\centering
\includegraphics[width=0.7\textwidth]{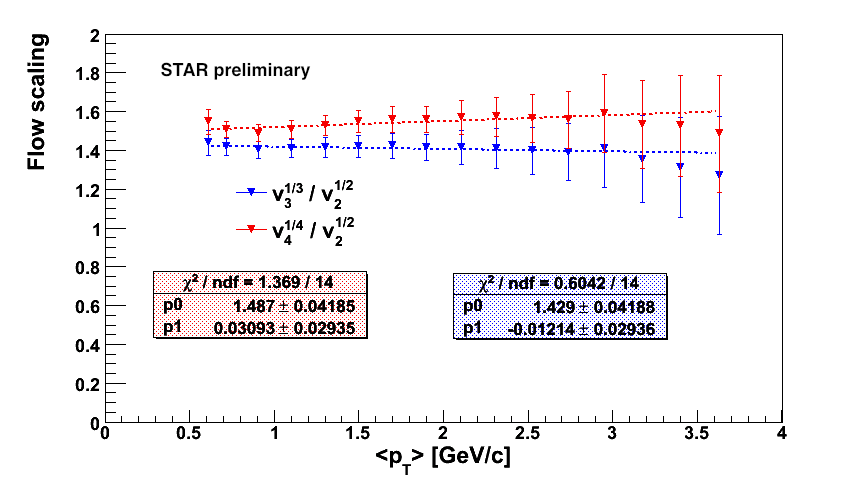}
\caption{Higher order harmonic scaling relations using, $v_{2}, v_{3}$ and $v_{4}$.}
\label{figure:seven}
\end{figure}

Finally we show scaling relations using higher order harmonics in order to compare our findings to certain hydro predictions [9]. Plotted in fig.~\ref{figure:seven} are the ratios $v_{4}^{1/4}/v_{2}^{1/2}$ and $v_{3}^{1/3}/v_{2}^{1/2}$. Theory predicts those ratios should be independent of $\langle p_{T}\rangle$ in the bulk $p_{T}$ regime. Even though we observe the predicted trends, in order to extract more information about the medium properties from the scaling ratios, we need to compare our results to theory applied to similar kinematic constraints (centrality and $p_{T}$) than the data.      

\subsection{Discussion}
We studied the evolution of di-hadron correlations in angular variables $\Delta\eta$, $\Delta\phi$ using Au+Au collisions at $\sqrt{S_{NN}}$ = 200 GeV. Our main focus was to study the long-range correlation observed in the central events bin by increasing the lower $p_{T}$ acceptance of both charged particles. By focusing on 0-10$\%$ centrality bin, we study the most violent heavy ion collisions where a formation of a quark gluon plasma is expected. Therefore our study of the long-range correlation could reveal further insight into the properties of the deconfined medium.\\
The measured structures evolve smoothly as a function of $\langle p_{T}\rangle$. The near-side correlation evolves from an elongated 2d Gaussian-like structure towards a structure which has a flat pedestal and a jet like peak on top of it. The away side structure broadens and flattens at higher $p_{T}$. Both the near side long-range correlation and away side amplitude reduces with increasing $p_{T}$. However, we observe a jet like peak strength growth (see fig.~\ref{figure:one}) after $\langle p_{T}\rangle$ $\approx$ 2.5 GeV/c possibly due to an increase of correlated pairs within a fragmenting high $p_{T}$ parton even though the total yield in the correlation function (including uncorrelated background) drops, as expected.\\
The data evolution was modeled using an empirical model function which has nine free parameters. We extract the contribution to the correlations due to initial energy density fluctuations via the use of higher order Fourier harmonics cos(n$\Delta\phi$), n = 1,2,3,4,5. In addition, the data require an asymmetric 2d Gaussian model component in order to fit the residual structure after subtracting the harmonics contribution.\\
The evidence for initial energy density fluctuations in the correlation data (see Fig.~\ref{figure:two}) implies that we do create a medium that behaves collectively. We further test the collectivity of the medium by testing our predictions against hydrodynamical scaling relations (see Fig.~\ref{figure:seven}) and observe the predicted trends. Theory states that any deviation of the scaling ratios from ideal hydro predictions could be due to partial thermalization effects [9]. However our present scaling data could not be directly compared to those predictions due to discrepancies in kinematic cuts. Nonetheless, our extracted $v_{n}$ parameters are useful in studying medium properties(e.g. viscosity and diffusion coefficients) and initial conditions (e.g. CGC vs. Glauber initial distributions) of the colliding heavy ions.\\
The remainder component after subtracting harmonics, can be viewed in two parts. Below $\langle p_{T}\rangle$ $\approx$ 2.5 GeV/c the remainder widths are asymmetric and the amplitude drops as a function of $\langle p_{T}\rangle$. In the same $\langle p_{T}\rangle$ region, the widths are modified compared to p+p data. The relative asymmetries as well as width modulation with respect to p+p data occur at different $\langle p_{T}\rangle$ values. Nonetheless, the remainder below $\langle p_{T}\rangle$ $\approx$ 2.5 GeV/c is elongated in $\Delta\eta$ and more modified than the $\Delta\phi$ width in comparison to p+p parameters. Therefore the cause of the asymmetric remainder component is open to interpretation and the measured modulation factors could be useful input for jet shape modification models. On the other hand, the symmetric remainder component above $\langle p_{T}\rangle$ $\approx$ 2.5 GeV/c shows a width modification of a factor two compared to p+p collisions. Theorists could use this information to constrain partonic energy loss models and calculation of energy transport coefficients in the deconfined medium.\\ 
In summary, we discussed that the near side long range correlation structure can be understood as arising from initial energy density fluctuations and possible modified jet phenomena. The modified asymmetric 2d Gaussian needs to be studied further in the context of jet fragmentation. Charge dependent and particle identified measurements are two suggested approaches for further studies.


\section*{References}
\begin{thebibliography}{9}
\bibitem{iopartnum} J. Adams et al. (STAR Collaboration), Phys. Rev. Lett. 95, 152301 (2005)
\bibitem{iopartnum} J.Putschke for the STAR collaboration, J. Phys. G34, S679 (2007)
\bibitem{iopartnum} G. Agakishiev et al. (STAR Collaboration), ÊarXiv:1109.4380
\bibitem{iopartnum} J. Porter and T. Trainor, J. Phys.: Conf. Ser. 27, 98 (2005)
\bibitem{iopartnum} S. Voloshin, Phys. Lett. B632, 490 (2006)
\bibitem{iopartnum} S. Gavin, G. Moschelli, L. McLerran, Phys. Rev. C79, 051902 (2009)
\bibitem{iopartnum} Xin-Nian Wang et al., Phys. Rev. Lett. 106, 162301(2011)
\bibitem{iopartnum} B. Alver et al., Phys. Rev. C 81, 054905 (2010)
\bibitem{iopartnum} C. Gombeaud et al., Phys. Rev. C 81, 014901 (2010) 
\end{thebibliography}
\end{document}